\documentclass{ws-ijmpd}

\begin{document}

\catchline{}{}{}{}{}
%
%%%%%%%%%%%%%%%%%%%%%%%%%%%%%%%%%%%%%%%%%%%%%%%%%%%%%%%%%%%%%%%%%%%%

\title{Mass-radius relation for magnetized strange quark stars}
%MANUSCRIPTS\footnote{For the title, try not to use more than 3 lines.
%Typeset the title in 10~pt Times roman, uppercase and boldface.}  }
%----------------------
\author{A. P\'{e}rez Mart\'{\i}nez}
%\email{aurora@icmf.inf.cu}
\address{Instituto de Cibern\'{e}tica, Matem\'{a}tica y F\'{\i}sica (ICIMAF), \\
Calle E esq 15 No. 309 Vedado, C. Habana, 10400, Cuba.}

\author{R. Gonz\'{a}lez Felipe}
%\email{gonzalez@cftp.ist.utl.pt}}
\address{Instituto Superior de Engenharia de Lisboa,\\
Rua Conselheiro Em\'{\i}dio Navarro, 1959-007 Lisboa, Portugal.}
\address{Centro de F\'{\i}sica Te\'{o}rica de Part\'{\i}culas, Instituto Superior T\'{e}cnico,\\
Avenida Rovisco Pais, 1049-001 Lisboa, Portugal.}

\author{D. Manreza Paret}
%\email{hugo@icmf.inf.cu}
\address{Facultad de F\'{\i}sica, Universidad de la Habana,\\
San L\'{a}zaro y L Vedado, C. Habana, Cuba.}

\maketitle

%\begin{history}
%\received{Day Month Year}
%\revised{Day Month Year}
%\comby{Managing Editor}
%\end{history}

\begin{abstract}
We review the stability of magnetized strange quark matter (MSQM) within the phenomenological MIT bag model, taking into account the variation of the relevant input parameters, namely, the strange quark mass, baryon density, magnetic field and bag parameter. A comparison with magnetized asymmetric quark matter in $\beta$-equilibrium as well as with strange quark matter (SQM) is presented. We obtain that the energy per baryon for MSQM decreases as the magnetic field increases, and its minimum value at vanishing pressure is lower than the value found for SQM, which implies that MSQM is more stable than non-magnetized SQM. The mass-radius relation for magnetized strange quark stars is also obtained in this framework.
\end{abstract}

\section{Introduction}
Strange quark matter (SQM)~\cite{weber} could be the ground state of matter if the conjecture of
Bodmer~\cite{Bodmer:1971we} and Witten~\cite{Witten:1984rs} is finally proved. If their hypothesis is true, this state could appear at the inner core of compact objects such as neutron stars or strange stars. Such a possibility has attracted a lot of attention in the astrophysical context since it could explain a set of observations which, presently, are not understood within the standard neutron star models~\cite{Lugones:2002va,Lugones:2002zd,Quan,Nice,Xu}. As SQM is more stable and bound at finite density, one expects configurations of quark stars with macroscopic properties quite different from neutron stars~\cite{itoh,Hansel}. These objects would
be self-bound and their masses would scale with the radius as $M\sim R^{3}$.

To seek the equation of state (EoS) of quark matter it is necessary to resort to phenomenological models. Among the most popular ones are the MIT bag model and the Nambu-Jona-Lasinio (NJL) model, both having advantages and disadvantages. While for the MIT bag model confinement is guaranteed by the bag parameter, the model is unable to reproduce the chiral symmetry breaking at zero density. On the other hand, the NJL model exhibits chiral symmetry breaking but does not explain confinement. Regardless of its disadvantages, the bag model turns out to be quite satisfactory not only in describing quark matter at finite density but also when applied to a real astrophysical scenario in the presence of strong magnetic fields. It is well known that pulsars, magnetars, neutron stars and the emission of intense sources of X-rays could be associated to sources with intense magnetic fields around $10^{13}-10^{15}$~G or even higher fields~\cite{duncan,kouve}. The collapse of a supernova shrinks the size of the star by a factor of $10^{5}$. Its magnetic field, which might be initially small ($\sim 100$~G), increases dramatically and could grow up to field values of the order of $10^{12}$~G, which could be then further amplified through a dynamo-like effect, reaching $10^{15}$~G in a magnetar. Due to flux conservation during the core collapse, at the inner core of the star the magnetic field could be even larger. The classical theoretical estimates based on the scalar virial theorem indicate that fields as high as $10^{18}$~G could be allowed~\cite{mag1,mag2}.

In Ref.~\cite{Felipe:2007vb}, the SQM properties were investigated in the presence of a strong magnetic field, taking into account $\beta$-equilibrium and the anomalous magnetic moments (AMM) of the quarks. It was found that the stability of the system requires a magnetic field value $B \lesssim 10^{18}$~G, in contrast to the bound $B \lesssim 10^{19}$~G, obtained when the AMM are not considered~\cite{Chakrabarty:1996te}. Here we aim at:
\begin{enumerate}
\item Obtaining the equation of state (EoS) of MSQM;
\item Proving that the energy per baryon $E/A$ for MSQM is lower than the one of SQM and magnetized asymmetric quark matter (neutrally-charged normal matter formed by quarks $u$ and $d$ in $\beta$-equilibrium with electrons);
\item Determining stable mass-radius configurations of magnetized strange quark stars (MSQS) through the solution of the Tolman-Oppenheimer-Volkoff (TOV) equations~\cite{TOV}.
\end{enumerate}

\section{SQM in the presence of a strong magnetic field}
\label{sec2}

For a degenerate MSQM, the energy density, pressure components and number density are given by the expressions~\cite{Felipe:2007vb,Felipe:2008cm}
\begin{equation}
\varepsilon=\Omega +\mu N =B\sum_i{\mathcal M}^0_{i}\sum_{\eta=\pm 1}\sum_{n}\left (
x_ip^{\eta}_{F,i}+h_{i}^{\eta\,2}\ln\frac{x_i+p_{F,i}^{\eta}}{h_{i}^{\eta}}\right
),\label{TQi}
\end{equation}
\begin{equation} \label{Ppar}
  P_{\parallel}=-\Omega =B\sum_i {\mathcal M}^0_{i}\sum_{\eta=\pm 1}\sum_n\left ( x_{i}p_{F,i}^{\eta} -
   h^{\eta\,2}_{i}\ln\frac{x_{i} +
 p^{\eta}_{F,i}}{h^{\eta}_{i}}\right ),\end{equation}
\begin{equation}\label{Pper}
 P_{\perp} = -\Omega -{\mathcal M} B=B\sum_i {\mathcal M}^0_{i}\sum_{\eta=\pm 1} \sum_{n}\left (2h_{i}^{\eta}\gamma_i^{\eta}\ln\frac{x_{i} + p_{F,i}^{\eta}}{h_{i}^{\eta}}\right),
\end{equation}
\begin{equation}
 N= \sum_i N^0_{i}\frac{B}{B^c_{i}}\sum_{\eta=\pm 1}\sum_{n} p^{\eta}_{F,i}\,,
\qquad {\mathcal M}^0_{i}= \frac{e_i d_i m_{i}^2}{4\pi ^2}, \qquad
N_{i}^0 =  \frac{d_i m_{i}^3}{2\pi^2}\,,\label{TQf}
\end{equation}
where
\begin{eqnarray} \label{dimvar}
x_{i}=\mu _{i}/m_{i}, \qquad   h_{i}^{\eta} =
\sqrt{\frac{B}{B^{c}_i}\, (2n + 1-\eta) +1} -\eta y_{i}B\,,\nonumber\\
p_{F,i}^{\eta} = \sqrt{x_{i}^2-h_{i}^{\eta}\;^2},\qquad
\gamma^{\eta}_i=\frac{B\,(2n+1-\eta)}{2B^c_i\sqrt{(2n+1-\eta)B/B^c_i+1}}-\eta
y_i B\,,
\end{eqnarray}
are dimensionless quantities. In Eqs.~(\ref{TQi})-(\ref{TQf}), the sum over the Landau level $n$ is up to the integer value $n_{max}^i = I\left[\left((x_{i} +  \eta y_iB)^2 -1\right)\,B^{c}_i/(2B)\right]$; $x_i$ is the dimensionless chemical potential; $p_{F,i}$ and $h_{i}^{\eta}$ correspond to the modified Fermi momentum and magnetic mass, respectively; $B^{c}_i=m_{i}^2/|e_i|$ is the critical magnetic field, $y_i=|Q_i|/m_i$ accounts for the AMM and $\eta=\pm 1$ correspond to the orientations of the particle magnetic moment, parallel or antiparallel to the magnetic field. Finally, $d_e=1$ and $d_{u,d,s}=3$ are degeneracy factors. Different expressions for the parallel $P_{\parallel}$ and transverse $P_{\perp}$ pressures reflect the anisotropy of pressures due to the magnetic field.
Notice also that the quantities defined above contain not only the contribution of Landau diamagnetism (quantization of Landau levels) but also Pauli paramagnetism due to the presence of quark AMM. Since the latter do not significantly modify the equation of state and stability region of MSQM, from now on we shall neglect their effect.

\section{Stability window for magnetized SQM}
\label{sec3}
Let us first study the stability of SQM in the presence of a strong magnetic field as a function of the relevant input parameters of the model, i.e. the baryon density $n_B$, the magnetic field $B$, the bag parameter $B_{\rm bag}$ and the quark mass $m_s$. Since in a strong magnetic field the anisotropy of pressures implies $P_{\perp} < P_{\parallel}\,$~\cite{Felipe:2007vb}, within the MIT bag framework the stability condition for quark matter is
\begin{equation} \label{stabpper}
P_{\perp} =\sum_{i}P_{\perp,i}- B_{\rm bag}=0\,.
\end{equation}
In order to obtain the EoS of MSQM as well as of the magnetized asymmetric quark matter in
the stellar scenario under consideration (i.e. the inner core of neutron stars or quark stars)
we should also take into account a set of equilibrium conditions: $\beta$-equilibrium, baryon
number conservation and electric charge neutrality. Lepton number is not conserved because
neutrinos are assumed to enter and leave the system freely ($\mu_\nu=0$). The conditions for
magnetized SQM are determined by the following relations: $\mu_u+\mu_e=\mu_d\,$,\,
$\mu_d=\mu_s, \, 2N_u-N_d-N_s-3N_e=0\,$,\, $N_u+N_d+N_s=3 n_{B}$. Considering magnetized
asymmetric quark matter in $\beta$-equilibrium simply corresponds to set $\mu_s=0$ in the
above equations. Solving this system of equations it is then possible to obtain the parameter
 region which verifies the stability inequalities:
\begin{equation}\label{stabineq}
\left.\frac{E}{A}\right|_{SQM}^B<\left.\frac{E}{A}\right|_{SQM}^{B=0}<\left.\frac{E}{A}\right|_{^{56}\text{Fe}}
<\left.\frac{E}{A}\right|_{u,d}^{B}<
\left.\frac{E}{A}\right|_{u,d}^{B=0}\,,
\end{equation}
where $\left. E/A\right|_{^{56}\text{Fe}} \simeq 930$~MeV is the energy per baryon of the
iron nuclei. As shown in Refs.~\cite{Farhi:1984qu}, the bound $B_{\rm bag}> 57$~MeV/fm$^3$
must be imposed so that under the same conditions, at $P=0$ and $T=0$, the energy per baryon
of normal matter (quark matter composed by $u$ and $d$ quarks) is higher than the one of
nuclear matter, i.e. $\left.E/A \right|_{u,d} > m_n$, where $m_n \simeq 939$~MeV is the
neutron mass.

\begin{figure}[t]
      \begin{minipage}[t]{0.45\linewidth}
      \centering
      \includegraphics[height=5cm,width=6cm]{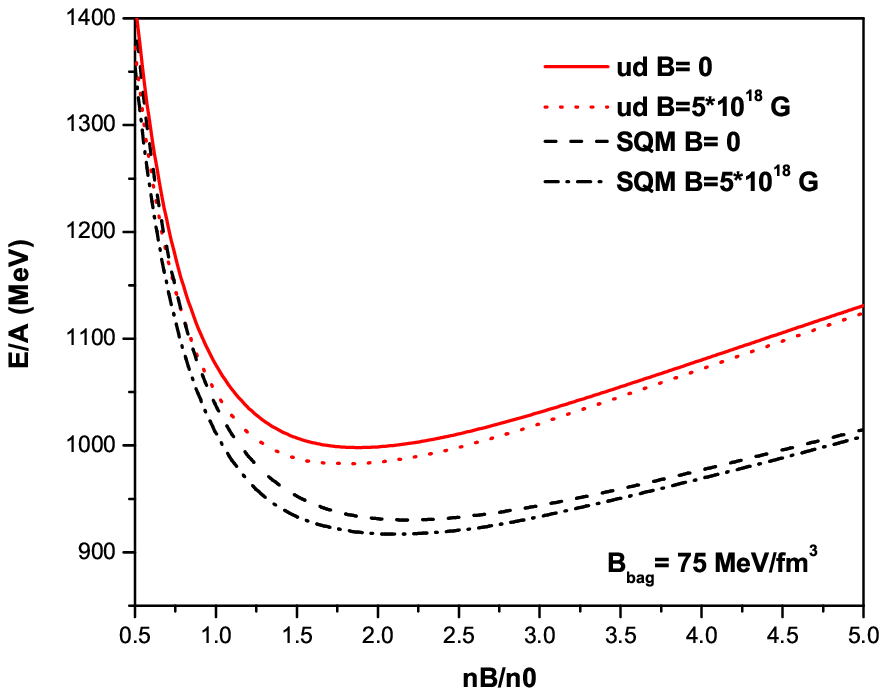}
      \caption{Energy per baryon as a function of the baryon density for $B=0$ and $B=5\times 10^{18}$~G, assuming $B_{\rm bag}=75$~MeV/fm$^3$.}\label{Energy}
      \end{minipage}
      \hspace{0.5cm}
      \begin{minipage}[t]{0.45\linewidth}
      \centering
      \includegraphics[height=5cm,width=6cm]{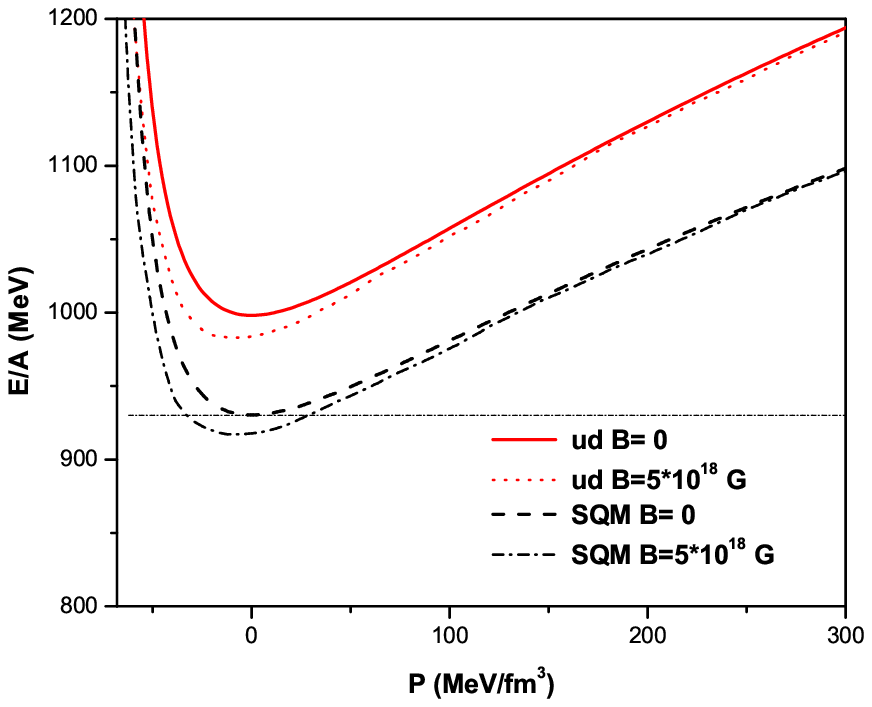}
      \caption{Energy per baryon versus pressure for $B=0$ and $B=5\times 10^{18}$~G. We take $B_{\rm bag}=75$~MeV/fm$^3$. The horizontal dot-dashed line corresponds to  $\left. E/A\right|_{^{56}\text{Fe}} \simeq 930$~MeV.}\label{EP}
      \end{minipage}
 \end{figure}

In Figure~\ref{Energy} we present a comparison of the energy per baryon $E/A$ (or, equivalently, $\varepsilon/n_{B}$) versus the number density $n_B/n_0$ ($n_0 \simeq 0.16$~fm$^{-3}$) in the absence of a magnetic field and for a magnetic field value of $5\times 10^{18}$~G. The bag parameter has been chosen to be $B_{\rm bag}=75$~MeV/fm$^3$. The plot also includes the asymmetric quark matter case. We assume $m_u = m_d = 5$~MeV and $m_s =150$~MeV in all cases. As can be seen from the figure, $E/A$ is always lower in the presence of a magnetic field. The corresponding behavior of $E/A$ with the pressure $P$ is shown in Figure \ref{EP}. One can notice that the point of zero pressure for MSQM is reached for an energy density value lower than that of SQM. Consequently, strange matter is indeed more stable and more bound in the presence of a magnetic field. Magnetized asymmetric quark matter has also a lower energy than the non-magnetized one, but higher than $^{56}\text{Fe}$ as expected.  At the zero-pressure point the corresponding baryon density is $n_{B} \simeq 2.18\, n_0$ for $B=0$ and $n_{B} \simeq 2.08\, n_0$ for $B=5\times 10^{18}$~G.

\begin{figure}[t]
%\begin{minipage}[t]{0.45\linewidth}
      \centering
      \includegraphics[height=5cm,width=6cm]{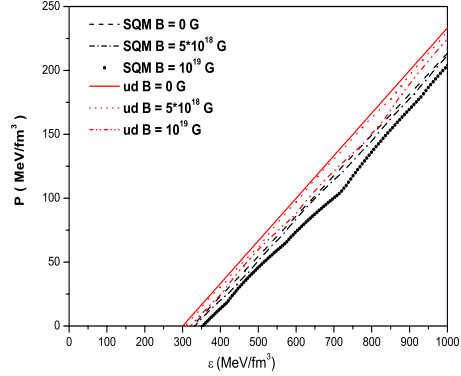}
      \caption{EoS: Pressure versus energy for $B=0$ and $B=5 \times 10^{18}$~G. For comparison,
      the EoS of normal quark matter in equilibrium with electrons is also plotted. }\label{EoS}
%      \end{minipage}
\end{figure}

To determine the EoS of MSQM, the system of equilibrium conditions must be solved numerically. In Figure~\ref{EoS}, we show the EoS for SQM (i.e. when $B=0$) and for MSQM for a magnetic field value of $5\times 10^{18}$~G. For comparison, we also include the EoS of normal quark matter in equilibrium with electrons. Although the changes in the EoS are not too significant, they can alter macroscopic observables such as the star mass and radius, as shall be seen in the next section.

\begin{figure}[t]
      \begin{minipage}[t]{0.45\linewidth}
      \centering
      \includegraphics[height=5cm,width=6cm]{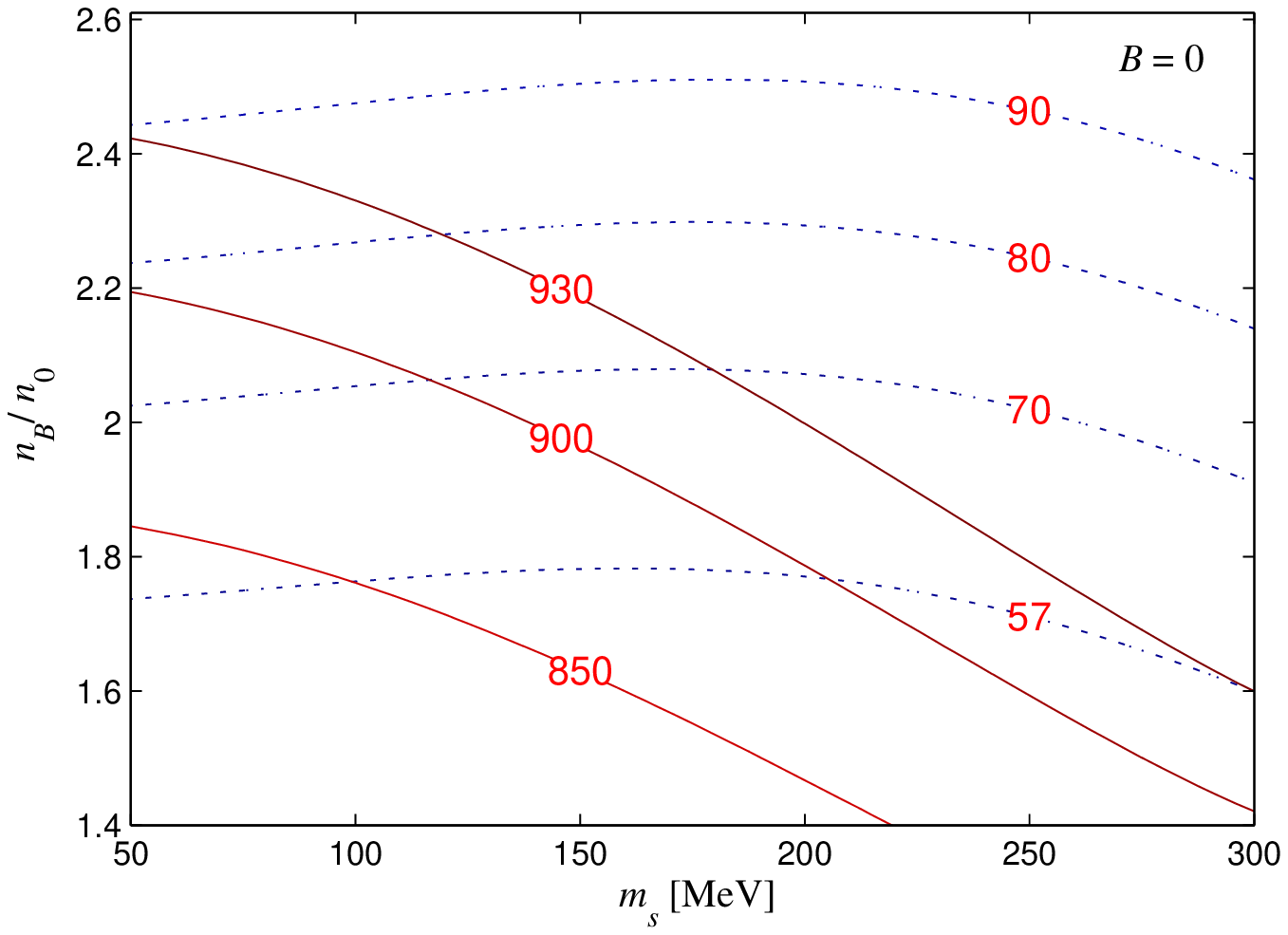}
      \caption{The stability windows of SQM in the ($m_s,n_B$)-plane. The
contours of constant $B_{\rm bag}$ (dashed lines) and $E/A$ (solid
lines) are shown.}\label{cBag}
      \end{minipage}%
      \hspace{0.6cm}
      \begin{minipage}[t]{0.45\linewidth}
      \centering
      \includegraphics[height=5cm,width=6cm]{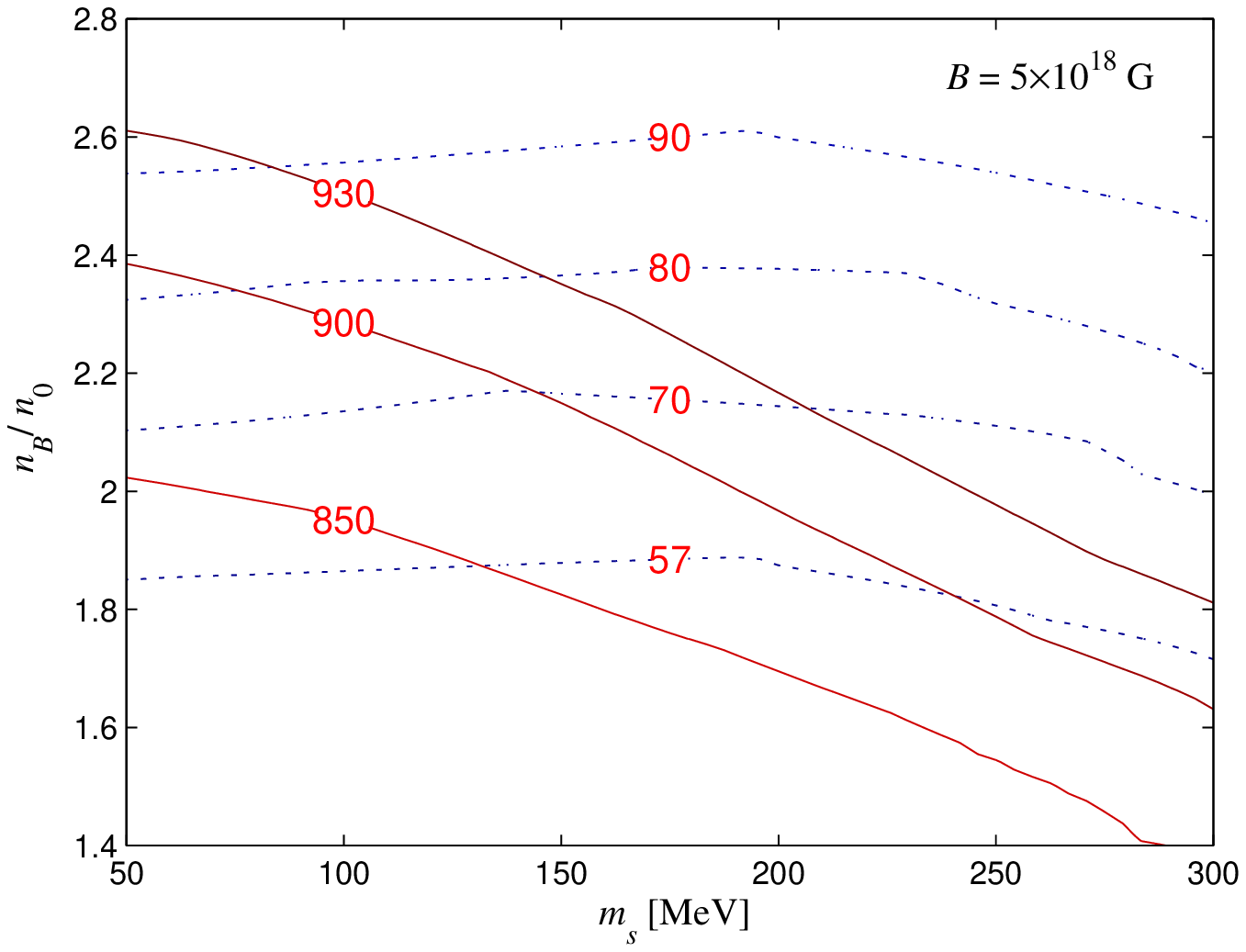}
      \caption{The stability windows of MSQM in the ($m_s,n_B$)-plane for
$B=5\times 10^{18}$~G. The contours of constant $B_{\rm bag}$
(dashed lines) and $E/A$ (solid lines) are shown.}\label{cE}
      \end{minipage}
 \end{figure}

Next we study the stability window for strange quark matter. In order to investigate how the magnetic field affects this window, we consider the stability regions in the ($m_s,n_B$)-plane. For illustration, we fix the magnetic field to $B=5 \times 10^{18}$~G and study the contours of $B_{\rm bag}$ and $E/A$. Such contours are presented in Figures~\ref{cBag} and \ref{cE} for SQM and MSQM, respectively. As can be seen from the figures, the magnetic field tends to shift the stability window of SQM towards higher values of the baryon density. It also modifies the allowed interval of $B_{\rm bag}$. Since below the energy contour of 930 MeV the EoS for MSQM corresponds to an $E/A$ at vanishing pressure lower than that of $^{56}$Fe, it can be considered absolutely stable.

\section{SQS mass-radius relation}
\label{sec4}

Configurations of spherical symmetric non-rotating compact stars are obtained by the numerical integration of the TOV equations~\cite{TOV} supplemented with the EoS. The radius $R$ and the corresponding mass $M$ of the star are determined by imposing the zero-pressure condition $P(R)=0$.  The EoS fixes the central pressure, $P(0)= P_c\,$, which together with the condition $M(0)=0$, completely determine the system of equations. It is worth remarking that despite the anisotropy of pressures in the presence of the magnetic field, this effect is not significant for $B
\lesssim 10^{19}$~G. Therefore, in our astrophysical context, even if the EoS is determined by the condition given in Eq.~ (\ref{stabpper}), TOV equations can be solved using a spherical metric~\cite{Herrera}.

\begin{figure}[t]
%
%      \hspace{0.6cm}
%      \begin{minipage}[t]{0.45\linewidth}
      \centering
      \includegraphics[height=5cm,width=6cm]{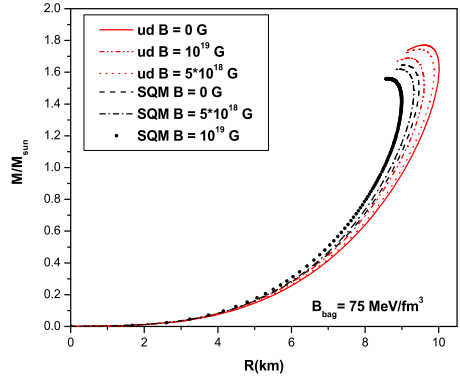}
      \caption{Possible $M-R$ configurations obtained for MSQS. The curves are shown for different values of the magnetic field $B$. For comparison, the $M-R$ relation of configurations of quark stars made of asymmetric quark matter are also shown. The most compact are the MSQS.}
\label{tov}
%      \end{minipage}
 \end{figure}

In Figure~\ref{tov} we plot the stable configurations of SQM without a magnetic field, as well as of MSQM and magnetized asymmetric quark matter for two different values of the magnetic field, $B=5\times 10^{18}$~G and $10^{19}$~G. The curves are presented for $B_{\rm bag} = 75$~MeV/fm$^3$. As already mentioned, the magnetic field enhances the stability of SQM and, therefore, it allows the appearance of stable stars with masses and radii smaller than those of non-magnetized SQM and magnetized asymmetric quark matter.

\section{Conclusions}
\label{sec5}
We have investigated the stability windows of magnetized strange quark matter within the phenomenological MIT bag model, taking into account the variation of the strange quark mass, the baryon density, the magnetic field and the bag parameter. Our conclusions can be summarized as follows:
\begin{enumerate}
\item The stability inequalities (\ref{stabineq}) are fulfilled for a wide range of the parameters of the bag model.

\item MSQM is indeed more stable than non-magnetized SQM and asymmetric quark matter. While for SQM the stability range for the baryon density is $1.8 \lesssim n_B/n_0 \lesssim 2.4$ for $50 \leq m_s \leq 300$~MeV, MSQM allows densities in the range $1.85 \lesssim n_B/n_0 \lesssim 2.6$ for $50 \leq m_s \leq 240$~MeV and a magnetic field value of $5\times 10^{18}$~G. Moreover, the allowed range for the bag parameter is $57 \lesssim B_{\rm bag} \lesssim 90$~MeV/fm$^3$.

\item The presence of strong magnetic fields at the inner core of compact stellar objects could lead to the existence of stable strange stars with smaller masses and radii.
\end{enumerate}

\section*{Acknowledgments}
We are grateful to J. Horvath for suggestions and useful comments. The work of A.P.M. has been supported by \emph{Ministerio de Ciencia, Tecnolog\'{\i}a y Medio Ambiente} under the grant CB0407 and the ICTP Office of External Activities through NET-35. A.P.M. also acknowledges TWAS-UNESCO for financial support at CBPF-Brazil as well as the PCI programme of CNPq Brazil. The work of R.G.F. was supported by \emph{Funda\c{c}\~{a}o para a Ci\^{e}ncia e a Tecnologia} (FCT, Portugal) through the project CFTP-FCT UNIT 777, which is partially funded through POCTI (FEDER).

\end{document}